\documentclass[12pt]{spieman}  
\usepackage{amsmath,amsfonts,amssymb}
\usepackage{graphicx}
\usepackage{setspace}
\usepackage{tocloft}

\title{Study of Vibrations at SOAR}

\author[a,*]{Andrei Tokovinin}
\affil[a]{Cerro Tololo Inter-American Observatory -- NSFs NOIRlab, Casilla 603, La Serena, Chile}

\cftpagenumbersoff{figure}
\cftpagenumbersoff{table} 
\begin{document} 
\maketitle

\begin{abstract}
The  4.1-m Southern  Astrophysics Research  (SOAR) telescope  in Chile
resembles larger  optical telescopes  by its thin  actively controlled
primary mirror,  built-in tip-tilt correction,  lightweight structure,
and adaptive  optics (AO)  turbulence correction.   The non-stationary
vibration of the  SOAR optical axis with  a frequency of 50  Hz and an
amplitude  reaching 20--30  mas has  been detected  by the  AO system,
degrading   the    quality   of   optical    speckle   interferometric
data. Accelerometers  revealed that  the 50\,Hz  tremor, driven  by an
external source, propagates within telescope structure and is strongly
amplified by  the response of  the fast tip-tilt mirror  servo system,
producing  characteristic elliptical beam  path. The AO  system also
evidenced periodic components  at 47\,Hz in the  defocus (matching the
frequency of fans in electronic  racks and computers) and at $\sim$65
Hz  in the  astigmatism.   The latter  is  associated with  structural
resonances of the  SOAR primary mirror support,  apparently excited by
the wind. Periodic tracking errors with typical frequencies of 0.5-2.5
Hz are mostly caused by periodic  errors of the encoders, depending on
their alignment,  and can be amplified  by the mount servo  at fast
tracking rates.   Vibration characterization  at SOAR  informs similar
studies at other telescopes.
\end{abstract}

\keywords{telescopes; adaptive optics; speckle interferometry}

{\noindent \footnotesize\textbf{*}  \linkable{andrei.tokovinin@gmail.com} 
 }


\section{Introduction}
\label{sec:intro}  

As  ground-based  telescopes  become  bigger, they  must  also  become
lighter to remain  within affordable budgets.  The cost  of a telescope
depends  on its  moving  mass,  so making  the  optics and  mechanical
structure lighter is a necessity. However, lighter telescopes are less
rigid and  more prone  to mechanical vibration.   The lack  of passive
mechanical  stiffness  is often  compensated  by  servo controls  that
effectively improve rigidity without  increasing the mass.  Artificial
rigidity is the mainstream in  modern telescope design.  On this path,
designers and users  of large telescopes have to  deal with vibrations
and their mitigation.

Almost all  large ground-based  telescopes are equipped  with adaptive
optics   (AO)  systems   for  partial   compensation  of   atmospheric
distortions (``seeing'') because high  angular resolution is essential
for reaching their full science  potential. However, vibrations of the
optical  axis  (jitter) and  other  wavefront  distortions inside  the
telescope  also  should  be  compensated   by  AO  together  with  the
seeing. Therefore,  vibrations are a relevant  factor in
the  design and  operation of  astronomical  AO systems.  Even a  tiny
jitter that  would be  harmless in seeing-limited  observations becomes
detrimental at high angular resolution ---  in the AO instruments and in
speckle interferometry.

\begin{figure}
\begin{center}
\begin{tabular}{c}
\includegraphics[width=10cm]{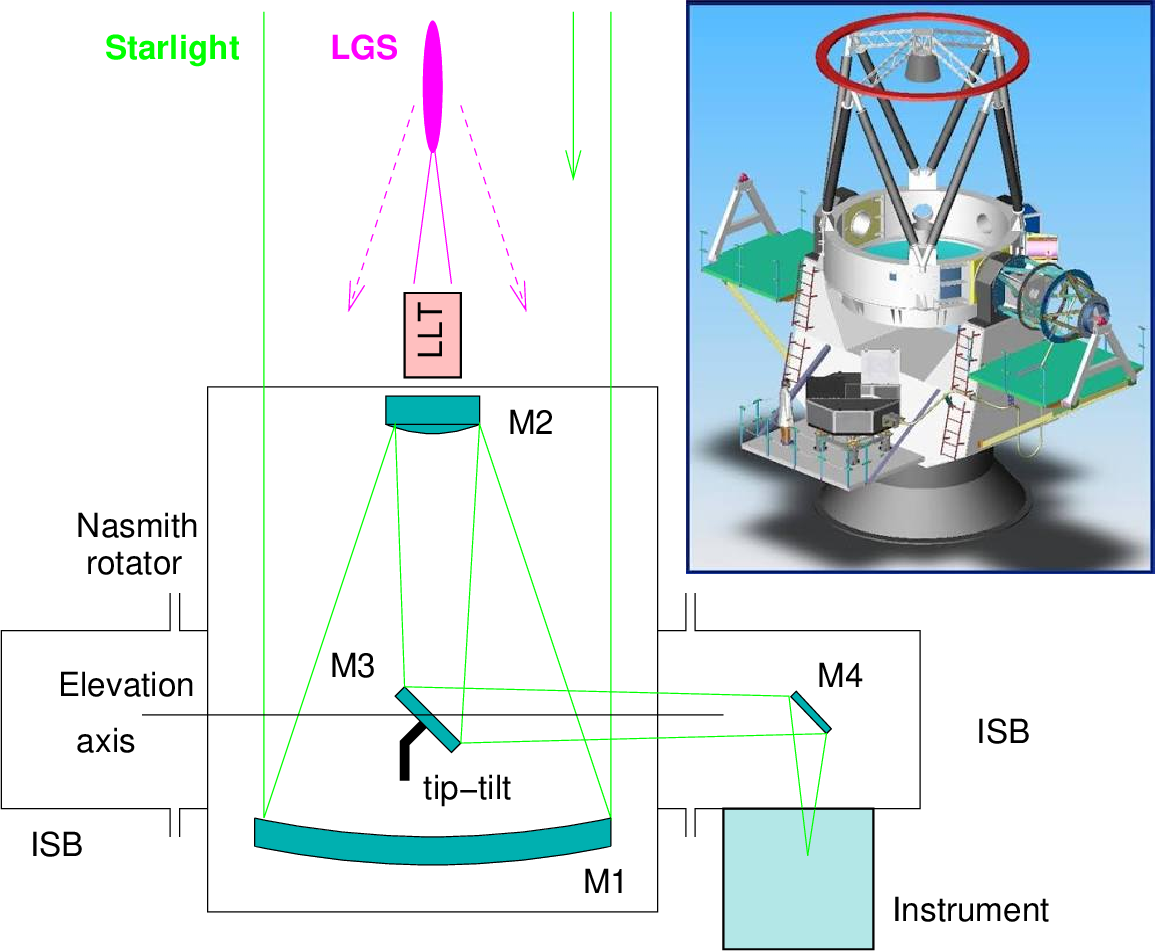}
\end{tabular}
\end{center}
\caption{Scheme of  the SOAR  telescope. Starlight  is focused  by the
  thin primary mirror  M1, reflected back by the  secondary mirror M2,
  and directed  along the elevation axis  by the actuated  tertiary mirror
  M3. At both  ends of the elevation axis, the  Nasmith rotators track
  the  on-sky  position angle  (the  mount  is alt-azimuth),  and  the
  instrument  selector boxes  (ISBs) switch  the beam  between science
  instruments using  mirrors or dichroics. The  laser guide star (LGS) is
  produced by  the small  laser-launch telescope (LLT)  located behind
  M2. The insert shows a 3D rendering of the SOAR  design.
\label{fig:SOAR} }
\end{figure}

This paper  considers vibrations  in the 4.1-m  SOuthern Astrophysical
Research (SOAR) telescope located in  Chile \cite{SOAR}.  Although by modern
standards the  aperture size is  modest, SOAR shares many  features of
larger  telescopes,  namely  the thin  actively  supported  monolithic
primary mirror  (like VLT  and Gemini  8-m telescopes),  fast tip-tilt
correction provided  within the  telescope itself (rather  than within
its instruments), and  an AO system, SAM \cite{SAM}. The  AO technical data
furnish  rich  information  on  the  wavefront  distortions of both  
atmospheric and  technical origin.  SOAR is   actively used for
diffraction-limited  imaging  by   speckle  interferometry.  From  the
outset,  SOAR aimed  at reaching  high angular  resolution at  visible
wavelengths, thus  complementing AO systems at  larger telescopes that
work mostly in the infrared. Main elements of SOAR relevant to this
paper are introduced in Fig.~\ref{fig:SOAR}.


Presence of a periodic 50-Hz vibration at SOAR has been noted
during its commissioning \cite{Warner2004}. It
was attributed originally  to the electric pick-up  noise and ignored.
However, in 2009, when SAM was  first tested with natural guide stars,
the  50-Hz  tip-tilt  jitter  of   the  optical  axis  became  obvious
\cite{SAM09}.  The  amplitude of this  vibration was variable  and, at
times, comparable to the SOAR diffraction limit (30\,mas). It degraded
the quality  of the speckle-interferometric  data. The source  of this
vibration remained  a mystery.   Several studies  using accelerometers
are reported below.  They confirmed the presence of a quasi-sinusoidal
mechanical vibration  with a  frequency of  exactly 50\,Hz  at various
locations within  the telescope, even  at its pier.  The  amplitude of
these oscillations, however, was orders  of magnitude smaller than the
optical jitter, and its character was different.  We surmised that the
50-Hz  vibration was  excited by  some source  outside the  telescope,
propagated    through   its    structure,   and    amplified.    Using
accelerometers, we explored empirically structural resonances in SOAR,
but none close to 50\,Hz was found. Finally, in 2018, during telescope
shutdown for mirror coating, we  found the amplification mechanism: it
was the fast tip-tilt tertiary mirror  (M3) and its servo control.  In
the following 2--3 years we noted a gradual decrease of this vibration
and,    as   a    consequence,   the    improved   quality    of   the
speckle-interferometric data.   The reason of this  positive trend has
not been identified.

This  paper reports  the  character  of the   SOAR vibrations  in
\S~\ref{sec:wf},  accelerometry  studies in  \S~\ref{sec:acc}, and
the study  of M3 in \S~\ref{sec:M3}.  For completeness, the
jitter  of the  SOAR  mount is  described  in \S~\ref{sec:jitter}.  In
\S~\ref{sec:sum} our  findings are  summarized and their  relevance to
other modern telescopes is discussed.

\section{Wavefront Jitter at SOAR}
\label{sec:wf}

\subsection{Wavefront Statistics from AO Data}
\label{sec:sam}  

\begin{figure}[ht]
\begin{center}
\begin{tabular}{c}
\includegraphics[width=14cm]{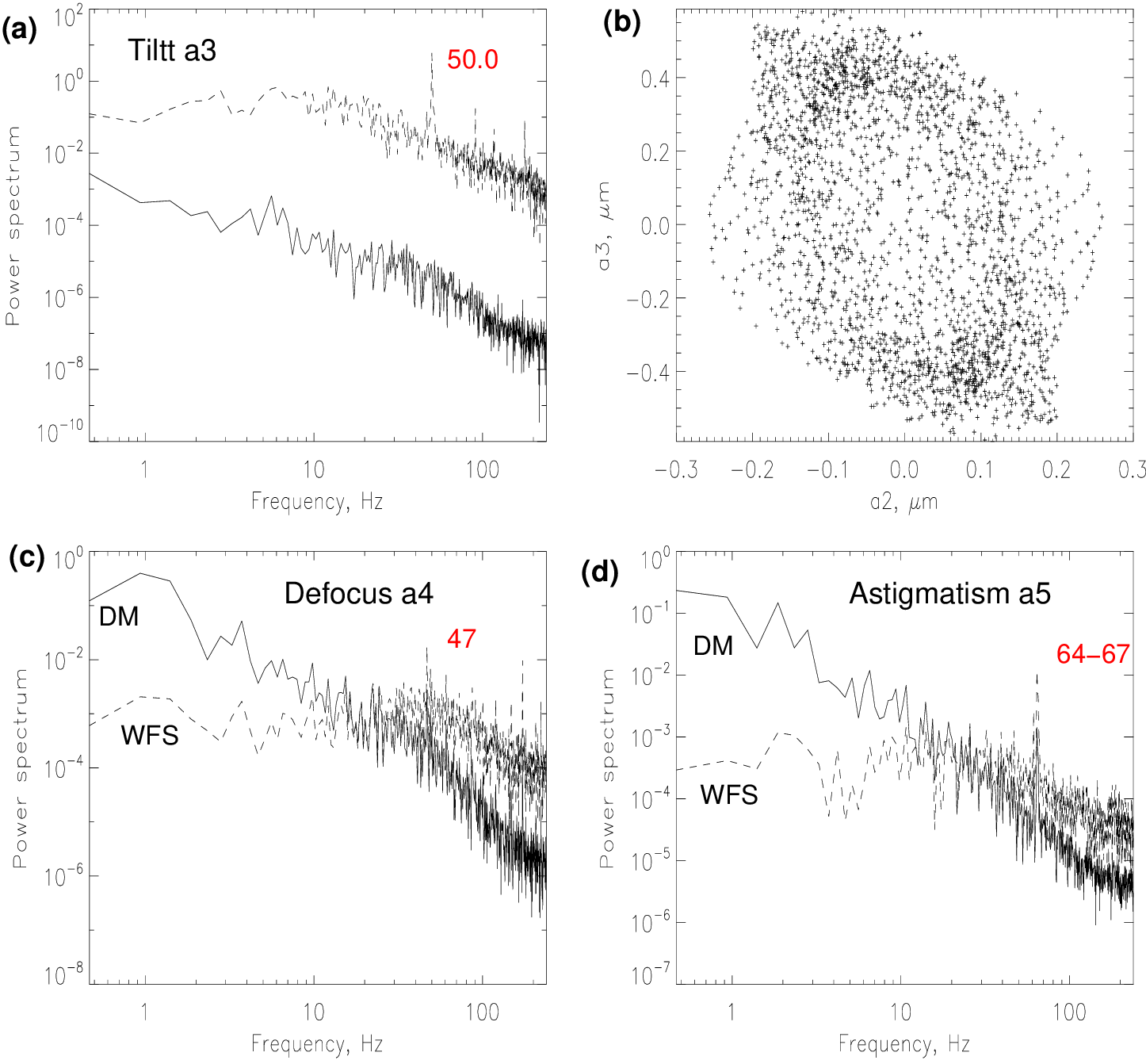}
\end{tabular}
\end{center}
\caption{Representative AO data taken on 2013-09-26 at UT 00:50. Panels
  (a), (c), (d) show  power spectra of the Zernike modes  3, 4, and 5,
  respectively, in  the log-log scale.  The dashed lines refer  to the
  WFS signal,  the solid lines to  the DM signal. Panel  (b) shows the
  trajectory  of  the  tilts  filtered in  the  48--52\,Hz  band.  The
  frequencies in Hz attributed to vibrations are marked in red.
 \label{fig:AO} }
\end{figure} 

The    SOAR     Adaptive    Module,     SAM,    is     presented    in
Ref.~\citenum{SAM}. High-order  wavefront distortions  are compensated
using an UV Rayleigh  laser guide star (LGS) and  a 10$\times$10 Shack-Hartmann
wavefront sensor (WFS). As the LGS  is located at a distance of 7\,km,
the WFS  senses predominantly the low-layer  turbulence.  The tip-tilt
errors are sensed by one or  two quad-cell guide probes located at the
input (uncorrected) focal  plane of SOAR, and compensated  by the fast
SOAR tip-tilt mirror, M3. In 2024, SAM has been upgraded by installing
a new 16$\times$16 WFS and a deformable mirror (DM) with a larger number of
actuators \cite{Faes2018,SAMplus}.

SAM  was  installed  at  SOAR  for the  first  time  in  2009  August.
Initially it  used the  natural guide  star \cite{SAM09}.  These tests
 revealed periodic  components in  the low-order  aberrations ---
tilts,  defocus, and  astigmatism.  Temporal  spectra of  higher-order
aberrations did not show any  periodic signals, only the atmospheric
distortions. This diagnostic was confirmed after 2011, when SAM
operated with the LGS.

During SAM commissioning and early observing runs, samples of the spot
coordinates in the WFS and the DM voltages were recorded
frequently. These binary records, referred to as {\em loop data,} were
explored by an IDL code that computed  Zernike coefficients
measured by the WFS and applied to the DM. A typical loop data record has a
length of 5\,s and contains over 2000 samples (the loop cycle was
2.089\,ms). The modal temporal power spectra were estimated by
splitting each record into 4 portions of 512 samples each and averaging
the resulting spectra in order to reduce the statistical noise. The
tip and tilt of the LGS are not compensated by the DM, and the
residual DM tilt is produced only by the  cross-talk involved in
the calculation of Zernike coefficients.  

Figure~\ref{fig:AO}  illustrates typical  signatures of  vibrations in
the SAM loop data. The 50-Hz  oscillations produce narrow peaks in the
Zernike tilts $a_2$ and $a_3$ in  most, but not all, records. Temporal
filtering  of  these signals  in  the  (48--52)\,Hz band  isolates  the
oscillations, revealing the phase shift between $a_2$ and $a_3$; thus,
the trajectory of  the optical axis oscillation is  elliptical. The parameters
of this ellipse vary and sometimes,  when the phase shift is small, it
resembles  a  tilted  line.   The  angular  displacement  $\alpha$  is
computed from the Noll's Zernike tilts in microns and the telescope
diameter $D$ as $\alpha = (4/D) a_{2,3}$, or
$0.2''$ per $\mu$m. So, the typical vibration amplitude of 0.2\,$\mu$m
corresponds to a  40\,mas wobble of the SOAR optical axis.

The defocus  term $a_4$  usually has  a component  at 47\,Hz,  and the
$45^\circ$ astigmatism  $a_5$ has a  broad feature around  65\,Hz. The
50-Hz component  is notable sometimes  in the coma  coefficients $a_7$
and $a_8$.

The signals  of the SAM  tip-tilt probes also have  the 50 Hz  line in
their temporal spectra. However, these quad-cell sensors do not have a
well-defined  response coefficient  between their  signal and  angular
displacement,  while   their  sampling  rate  of   100\,Hz  is  barely
sufficient (to see  the 50 Hz line clearly, we  interpolate the signal
to 200-Hz sampling before computing its power spectrum).

\subsection{Speckle Interferometry}
\label{sec:specjle}  

\begin{figure}[ht]
\begin{center}
\begin{tabular}{c}
\includegraphics[width=14cm]{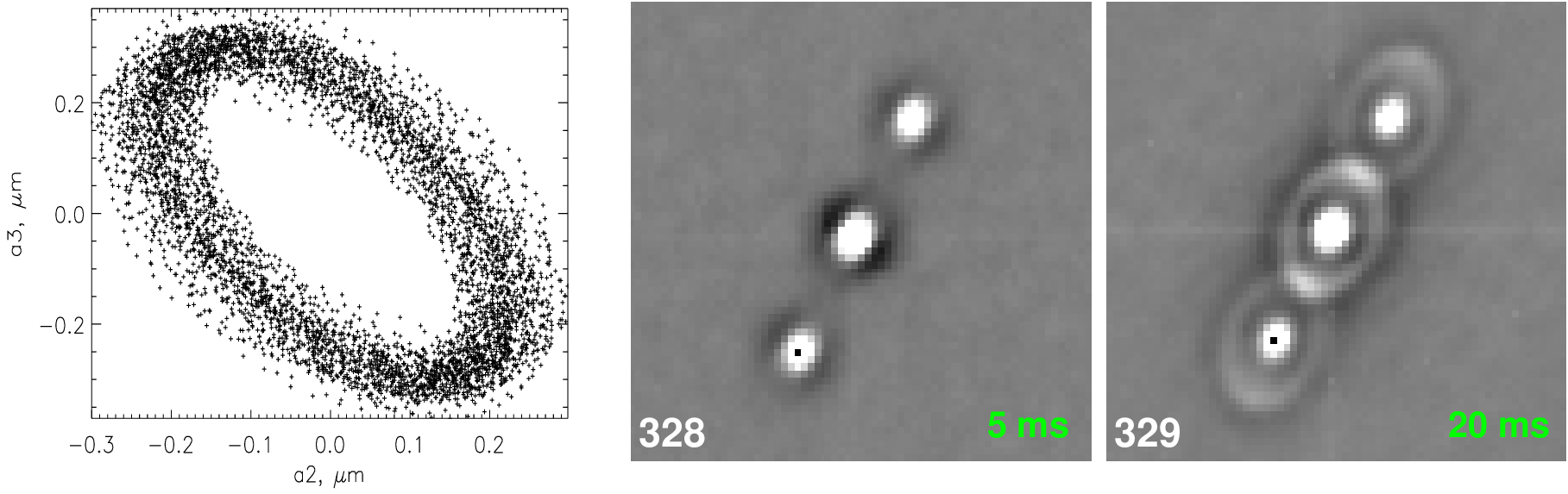}
\end{tabular}
\end{center}
\caption{Influence of the 50-Hz vibrations in speckle interferometry.  Left:
  trajectory  of the  line-of-sight  motion with  50\,Hz frequency,  as
  recorded by  SAM in closed loop on 2013-02-26 at UT 0:44.
Right: autocorrelations of  the speckle images of the  double star
HU~302 (separation $0.288''$) taken
on 2013-02-16 with exposure time of  5\,ms and 20\,ms. A short exposure
removes most  of the speckle  blur caused by  vibrations, at a  cost of
reduced magnitude limit.  The blur creates elliptical ``disks'' around
the stars.
 \label{fig:evidence} }
\end{figure} 

Speckle interferometry has been started at SOAR in 2007 as a technical
experiment  in  support  of  SAM  \cite{Cantarutti2008},  but  quickly
evolved into  an active science program  \cite{TMH10}.  The high-resolution
speckle camera,  HRCam, is  attached to  the visitor  port of  SAM. It
takes sequences of images (data cubes) with a fine spatial sampling of
15\,mas per pixel and short exposure times from a few to a few tens of
ms \cite{HRCAM}. Usually the AO  loop is not closed during speckle runs
for  the sake  of  efficiency and  flexibility, although  occasionally
faint stars were observed with HRCam and SAM in closed loop.

Vibration of  the SOAR optical  axis at  50\,Hz with an  rms amplitude
of $\sim$ 20\,mas   (peak-to  peak  image  displacement    60\,mas)
dramatically affects speckle interferometry  by reducing the signal at
high spatial  frequencies. A typical  exposure time of  20\,ms samples
one full  period of  the oscillations,  resulting in  a substantial
blur of the speckles. The blur shape  reflects the  axis trajectory: it  can be
nearly circular, creating  a ring-like halo (Fig.~\ref{fig:evidence}),
or  quasi-linear.  We  took  data at  different  position  angles  and
verified  that   the  blur   originates  upstream  from   the  SAM+HRCam
instrument.  Taking speckle  images with  exposures time  of 5\,ms  or
shorter   reduces  the   effect   of  vibration,   but also   the
sensitivity. The  non-stationary nature  of the 50-Hz  vibration means
that some (lucky) speckle data are degraded less. On the other hand,
non-stationarity prevents accurate calibration of the blur with
single reference stars observed before or after the science
targets. As noted above, the 50-Hz vibrations disappeared after 2020.

\section{Accelerometry and Mechanical Resonances}
\label{sec:acc}  


\subsection{Measurement Equipment}
\label{sec:hardware}  

Two accelerometers have been installed at SOAR by M.~Warner to monitor
the  behavior  of  its  structure  in  normal  operation  and  during
earthquakes, which  happen regularly in  this region. The  sensors are
triaxial accelerometers LI344ALH. Their  analog signals (voltages) are
acquired  continuously  with a  sampling  rate  of 200\,Hz  using  the
multi-channel digital-to-analog converter (DAC) model NI USB-6215. The
channels  0,  1,  2  correspond  to  the (X,  Y,  Z)  signals  of  the
accelerometer installed at  the SOAR top end near M2,  the channels 3,
4, 5  are (X, Y, Z)  signals of the second  accelerometer installed at
the   base   of   the   telescope   pier.    The   sensor   noise   is
50\,$\mu$gal/$\sqrt{\rm  Hz}$, and  the response  is 0.65  V/gal.  The
broad-band noise is about  200\,$\mu$gal rms.  Before acquisition, the
signal is  filtered to pass the  bandwidth of 150\,Hz.  With  a 200-Hz
sampling, this filter can pass some aliased high frequencies.

The data acquisition  computer runs under Windows  XP. The acquisition
software is  a free  version of the  LabView Signal  Express delivered
together with  the DAC.  It stores  the data  in a  proprietary binary
format. The data files are typically  accumulated for a few days, then
a new file is opened manually. The data, exported into text files, are
explored here using custom software written in IDL.

For a  harmonic motion with frequency $\nu$  and   amplitude
$x$, the acceleration amplitude  is $a = (2 \pi \nu)^2  x$.  If $a$ is
expressed in  gals (1 gal =  9.80665 m\,s$^{-2}$) and $\nu  = 50$\,Hz,
the amplitude in meters is 
\begin{equation}
x/a = 9.80655 (2 \pi \nu)^{-2}  = 9.363\,10^{-5} \;\; {\rm m/gal}.
\label{eq:x}
\end{equation}
A 50\,Hz  vibration with 1\,$\mu$m amplitude  corresponds to 0.01\,gal
or 6.5\,mV voltage, considering the response of 0.65 V/gal.

\subsection{Characteristics of the 50 Hz Acceleration}
\label{sec:50hz}  

We explored signal of the  accelerometer installed near SOAR M2 during
three  nights of  September  25-27,  2013, when  SOAR  was in  regular
operation. The  telescope movement in  elevation slowly changes  the Y
and Z  voltages due  to the varying  gravity direction.   These trends
were subtracted,  and the  power spectra of  small data  segments were
computed.

\begin{figure}[ht]
\begin{center}
\begin{tabular}{c}
\includegraphics[width=8.5cm]{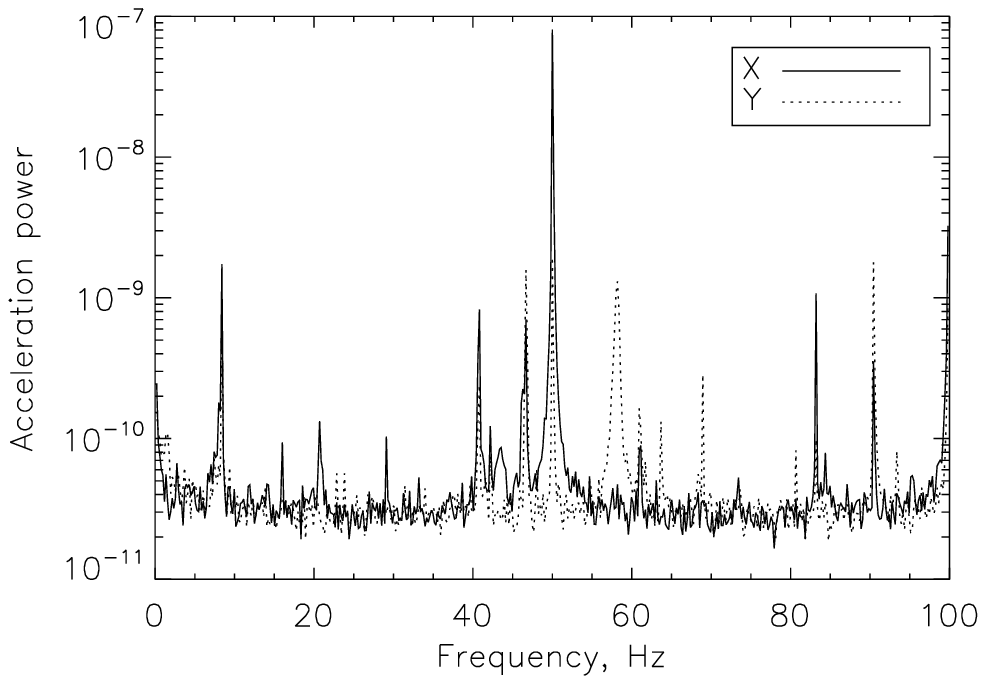}
\includegraphics[width=8.5cm]{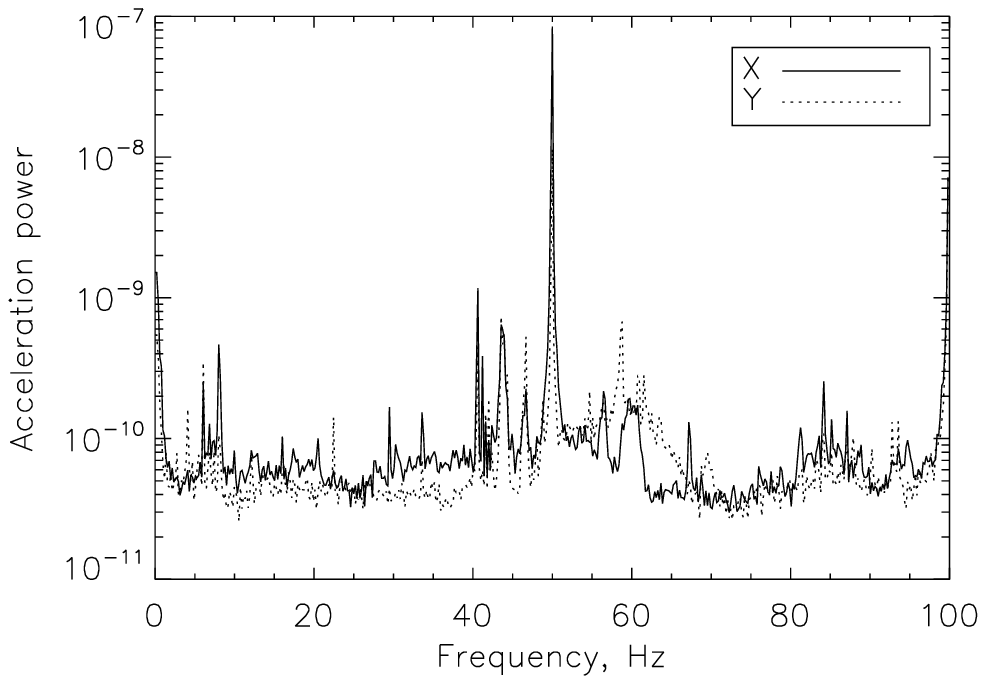}
\end{tabular}
\end{center}
\caption{Power spectra of acceleration at M2 recorded around UT 4h on 2013-09-25
  (left) and 2013-09-27 (right). Solid line --  along X, dashed line --
  along Y.
\label{fig:sep2013} }
\end{figure} 

Power spectra of the accceleration in  X and Y directions 
(perpendicular to the telescope axis) near M2  are shown in
Fig.~\ref{fig:sep2013} for  representative periods on two  nights. The
most prominent line is  at 50\,Hz. It is stronger in X  than in Y, and
is also present in Z. On September~25  we saw a line around 8\,Hz
(which could be an alias) and  some weaker lines.  The 47\,Hz line is
often seen in the  defocus term of SAM, and we note  its presence in the
acceleration as well. Different  data portions have different spectra,
but the 50-Hz line is present almost always. 

It  is known  that the  motion  of the  SOAR optical  axis with  50-Hz
frequency  in   X  and   Y  is  correlated   and  its   trajectory  is
elliptical. However, the two components  of the acceleration do not show
such  correlation, and the (X,Y) trajectory resembles a scatter plot. 
The lack of correlation proves  that the 50-Hz accelerometer signal is
not an  pickup noise and that it is not proportional to the
motion of the optical axis.

\begin{figure}[ht]
\begin{center}
\begin{tabular}{c}
\includegraphics[width=10cm]{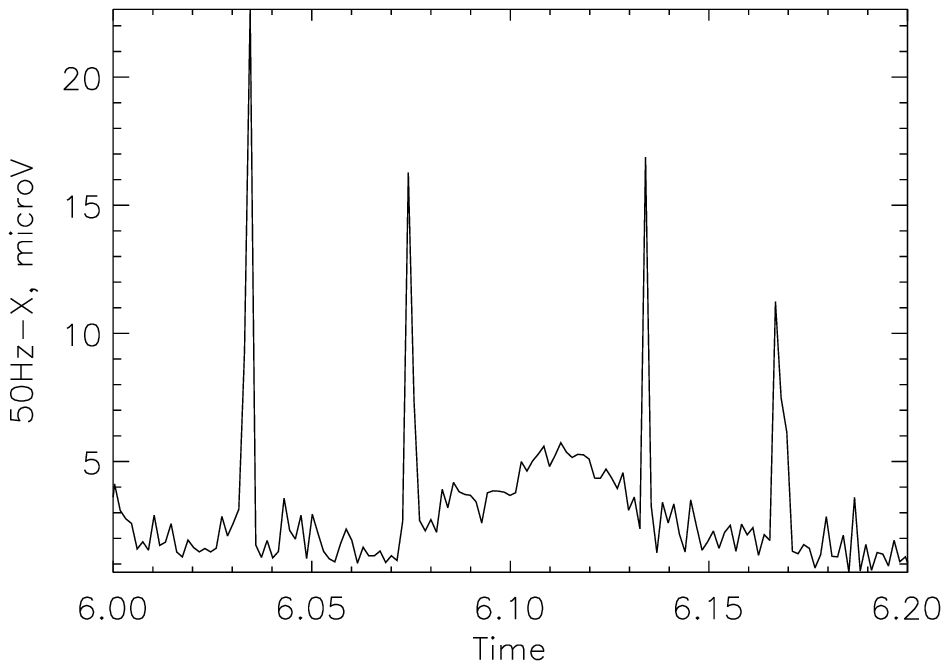}
\end{tabular}
\end{center}
\caption{Amplitude of the 50-Hz acceleration in X at M2 
   vs. UT time  on 2013-09-25. The spikes
   occur when the telescope slews to a new target. 
\label{fig:ampl-time} }
\end{figure} 

The amplitude  of the  50-Hz acceleration  changes with  time, without
obvious          dependence         on          the         telescope
position. Figure~\ref{fig:ampl-time} shows the  amplitude of the 50-Hz
acceleration in X during 12 minutes when the telescope pointed several
targets.  Between  the spikes that  coincide with slews,  the amplitude
does not remain stable, varying between 1 and 5 $\mu$V.

\begin{figure}[ht]
\begin{center}
\begin{tabular}{c}
\includegraphics[width=8.5cm]{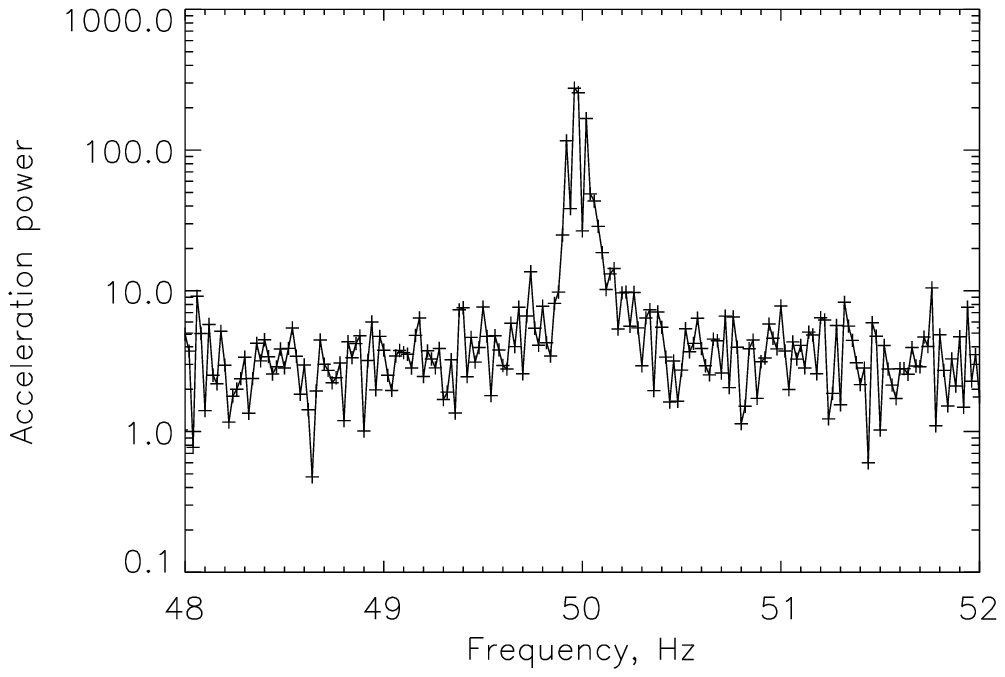}
\includegraphics[width=8.5cm]{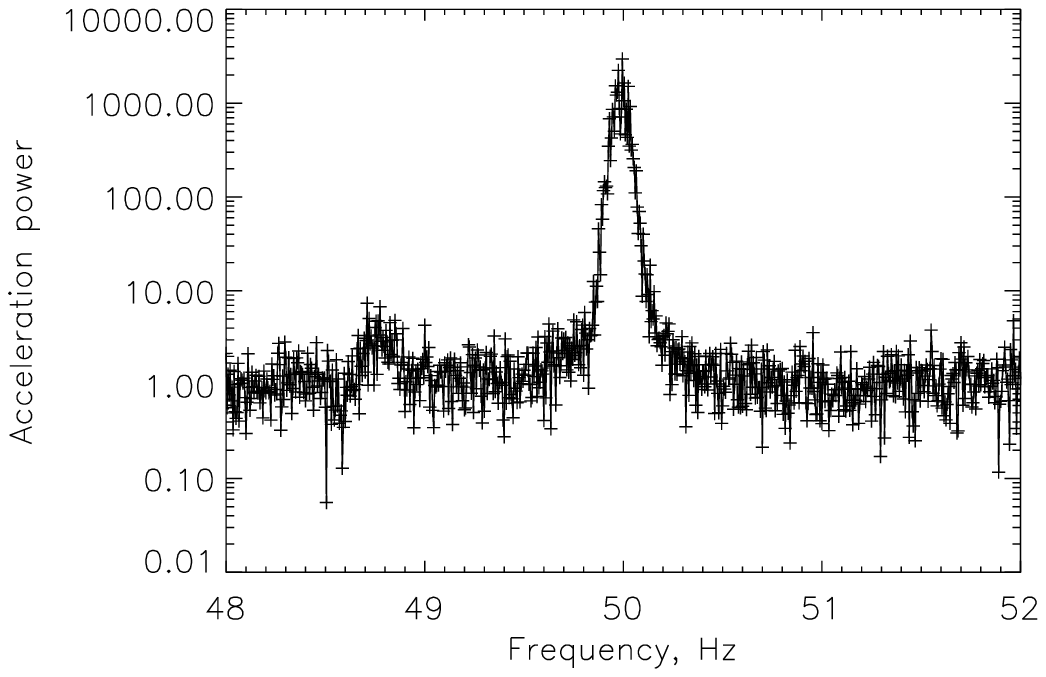}
\end{tabular}
\end{center}
\caption{Fragments of high-resolution temporal spectra of acceleration
  near 50 Hz. 
Left:  acceleration at  the SOAR  pier in Y on  2013-09-27 starting at 4h UT.
Right: acceleration at M2 on 2013-09-27 starting at 9.8h UT
when the telescope was parked. 
\label{fig:pier} }
\end{figure} 

The  acceleration at  the pier  of SOAR  was examined  using the  same
tools. Long data  segments allow us to measure the  width of the 50-Hz
line which is inversely proportional to the temporal coherence of this
signal. The half-width  of the line is about 0.25\,Hz  both at M2
 and at the pier base (Fig.~\ref{fig:pier}). The line width remains the
same  when  the  telescope  is  tracking and  when  it  is  stationary
(parked). The bandpass-filtered signal at  50 Hz shows oscillations in
each   coordinate   with   variable  amplitude   resembling   chaotic
interference (beating) between two close frequencies.

As noted above,  the amplitude of the 50 Hz  acceleration changes with
time, alternating between ``calm'' and ``noisy'' periods.  Analysis of
the amplitude during several hours did not reveal any regular patterns
or clues.  There  is only a very loose  correlation between amplitudes
of the 50-Hz acceleration at M2 and  at the pier; the amplitude at the
pier is approximately two times less than at M2.

If  SOAR vibrated  at 50\,Hz  as  a solid  body with  an amplitude  of
1\,$\mu$m (6.5\,mV acceleration)  at the top end, its  axis would move
with  an  angular   amplitude  of  45\,mas  (tube   length  of  4.5\,m
assumed). However,  as can be inferred  from Fig.~\ref{fig:ampl-time},
the actual  worst-case amplitude of the  50-Hz transverse acceleration
at M2  is $\sim$5\,$\mu$V, three  orders of magnitude smaller.  So, if
the  50-Hz acceleration  is indeed  the source  of the  50-Hz tip-tilt
vibration, it should be strongly amplified somewhere.


\subsection{Structural Resonances}
\label{sec:res}  

Pursuing the idea  that the 50 Hz vibration is  excited externally and
somehow amplified  within the telescope  structure, I  explored mechanical
resonances at SOAR.  On October  5, 2014, a portable accelerometer was
placed at  various locations on  the Instrument Selector Box  (ISB) at
the optical  Nasmith focus  of SOAR  to probe  whether it  vibrates or
not. The 50-Hz line was absent in  the records, instead we saw a 47-Hz
signal presumably  matching the  frequency of  fans in  the electronic
racks (asynchronous  motors). However,  these tests were  done without
telescope tracking.

\begin{figure}
\begin{center}
\begin{tabular}{c}
\includegraphics[width=12cm]{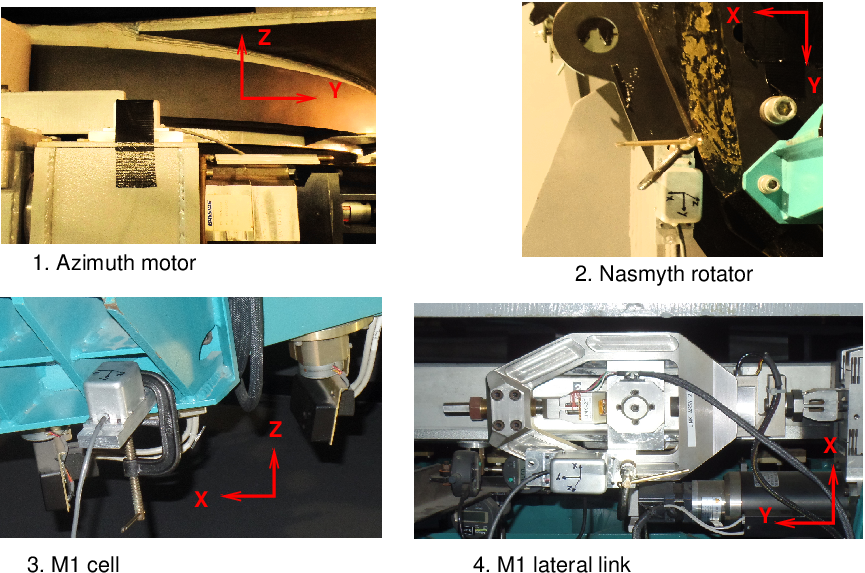}
\end{tabular}
\end{center}
\caption{Locations of the accelerometer during vibration tests on 2015-09-25.
\label{fig:locations} }
\end{figure} 

\begin{table}[ht]
\center
\caption{Accelerometer locations and  orientation on 2015-09-25}
\label{tab:loc}
\medskip
\begin{tabular}{c l ccc l}
\hline
$N$ & Location & X  & Y  & Z & Frequencies \\
\hline
1 & Azimuth motor   & Rad & Az & Up    & 50 \\
2 & Nasmyth rotator & Az  & Down & Rad &  (20-60); 50 \\
3 & M1 cell         & Az  & Rad & Up   & (20-30); 50  \\
4 & M1 link         & Up  & Az  & Rad  & (60-80); 66 \\
5 & Pier            & Down & Az & Rad  & (20); 50  \\  
\hline
\end{tabular}
\end{table}

\begin{figure}
\begin{center}
\begin{tabular}{c}
\includegraphics[width=8.5cm]{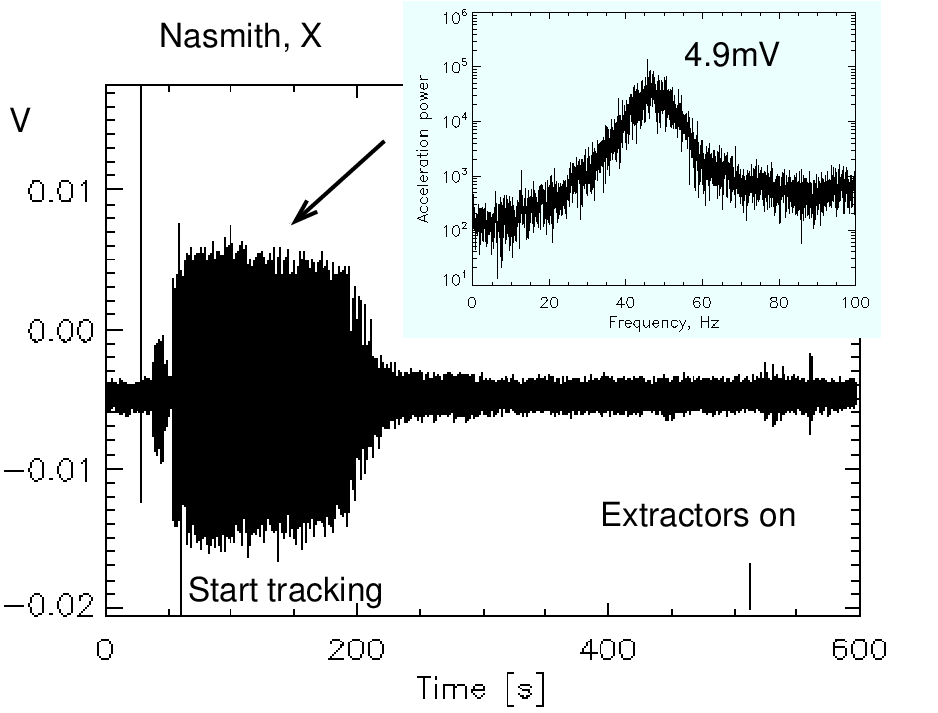}
\includegraphics[width=8.5cm]{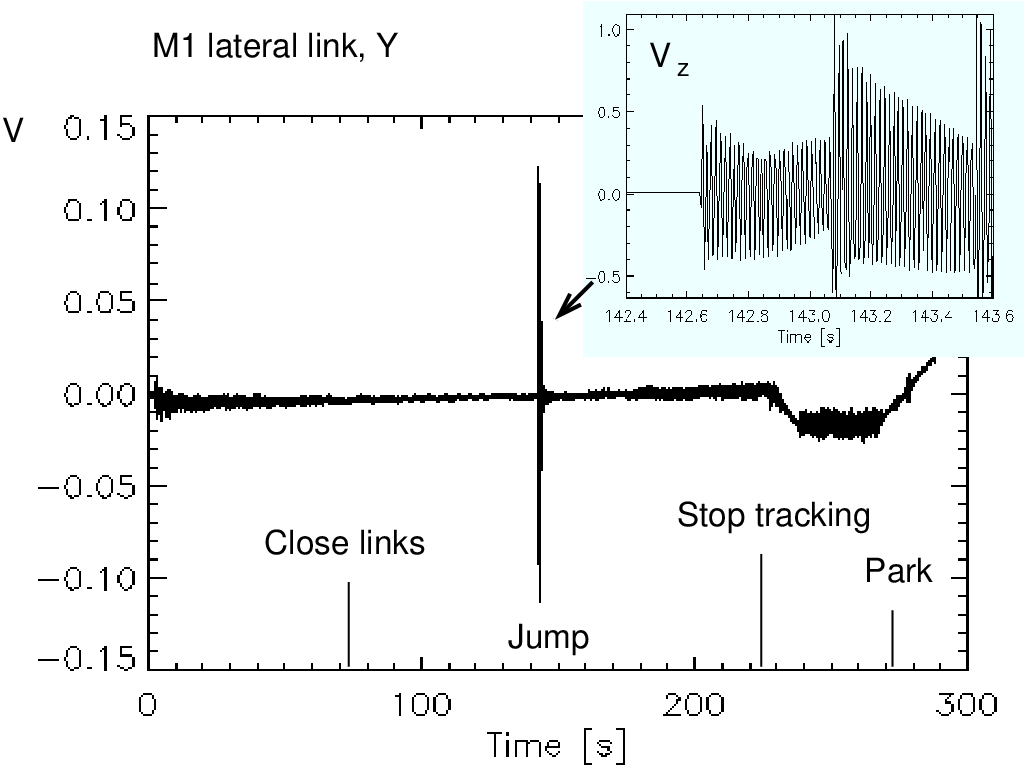}
\end{tabular}
\end{center}
\caption{Accelerometry  on  2015-09-25.   Left:  acceleration  at  the
  Nasmith  rotator along  X  with temporal  spectrum during the vibration
  episode. Right: Acceleration  at M1 lateral link in Y  with an artificial
  earthquake; the insert shows Z-oscillations during the quake.
\label{fig:vibtest} }
\end{figure}


\begin{figure}
\begin{center}
\begin{tabular}{c}
\includegraphics[width=12cm]{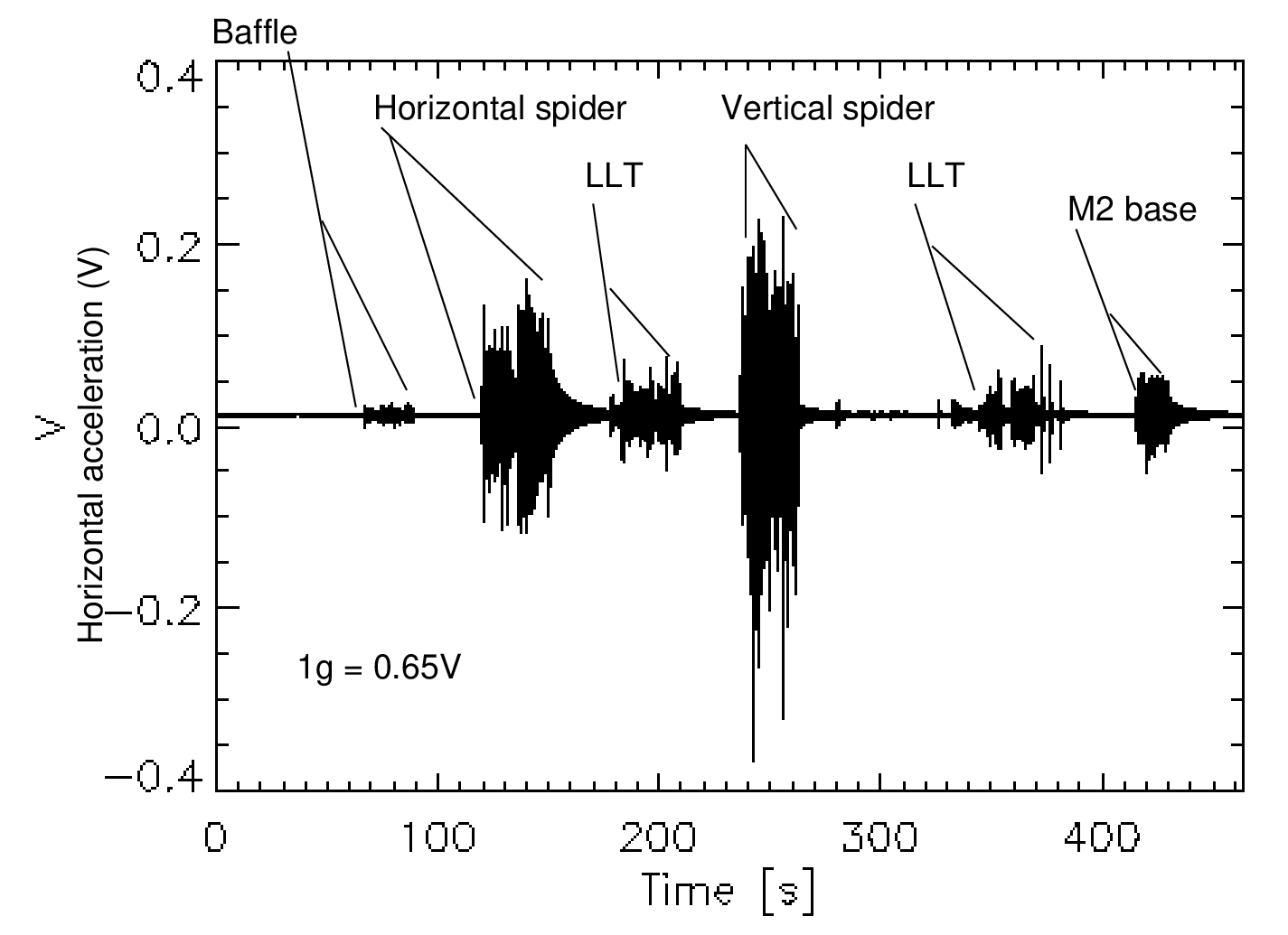}
\end{tabular}
\end{center}
\caption{Full record of the X-acceleration at the SOAR M2 taken
  on 2015-11-12. Moments when the structure was tapped are indicated.
\label{fig:top} }
\end{figure}

On September  25, 2015, we  studied the effect of  telescope tracking.
The same portable accelerometer (normally installed at the base of the
pier) was connected  by a long cable and placed  at various locations.
Simultaneous data  from the second accelerometer  installed behind the
M2 were recorded as  well.  Figure~\ref{fig:locations} shows the first
four tested locations. Table~\ref{tab:loc} lists the locations and the
directions of the accelerometer axes (Rad is parallel to the telescope
optical axis, Az is directed along azimuth).  At each location, one or
more     data    records     of    3     min.    or     longer    were
taken.   Figure~\ref{fig:vibtest}   illustrates   the    results
summarized below.

At the  azimuth motor (location  1), the 50-Hz  peak was seen  in the
signals of  all axes, with the  telescope  tracking or  not. The
motor  itself is  therefore not  responsible  for exciting  the 50  Hz
vibration. The rms  signal was $\sim$0.6\,mV.  In the  second record taken
at the same location  (with telescope moving in elevation) the  50 Hz line
was absent,  but weak lines at  47\,Hz and 83\,Hz appeared,  while the
rms signal in X and Y was smaller, 0.25\,mV. 

In the 10-min. record taken at the Nasmith rotator (location 2), there
is a  100-s period with  strong vibration (4.9 mV  rms in X,  i.e.  in
azimuth),  see  Fig.~\ref{fig:vibtest}  left.    It  began  after  the
telescope started tracking;  its spectrum had a broad  peak between 30
and 60 Hz.   The increased vibration could be caused  by some motor or
compressor that  work intermittently.  When it  stopped, the telescope
was still tracking, but the rms  signal dropped to 0.45\,mV, with only
a weak line at 50\,Hz.  At the  end of the record, the fans extracting
air   from  the   dome  were   turned  on,   without  any   effect  on
vibration.  During  the vibration  episode,  the  accelerometer at  M2
increased  the rms  only  by a  factor  of  two, not  by  an order  of
magnitude as at the Nasmith rotator.

The accelerometer attached to the M1 cell (location 3) recorded
relatively small accelerations (1.4, 0.4, and 0.6 mV rms at X, Y, Z,
respectively) independently of the telescope tracking. However, this
record also contains a ``vibration episode'' lasting about a
minute. During this period, the largest acceleration was in Z, and its
spectrum had a broad peak at 20--30 Hz. During the rest of the record,
only a narrow 50-Hz line was seen.  

Acceleration  at  the  M1  lateral   link  (location  4)  revealed  an
interesting phenomenon. The amplitude  was relatively small (1.1, 1.6,
1.5 mV  rms in X, Y,  and Z), and it  did not increase when  the servo
loop of  the lateral links  was closed.  A  narrow line at  66\,Hz was
present only in the X-axis (vertical).  In the middle of the record, people
standing  on the  Nasmith  platform jumped,  emulating an  earthquake.
This produced  large accelerations  in X and  Z (perpendicular  to the
link axis).  The  acceleration amplitude in Z exceeded 1  gal, and its
power  concentrated   between  60  and  80   Hz,  indicating  resonant
oscillations. The maximum  at $\sim$66\,Hz matches the broad  peak in the
astigmatism temporal spectra detected with SAM. This vibration is
therefore associated with the structural resonances of the M1 and its
support. During normal telescope operation, these resonances can be
excited by the wind buffeting. Quite surprisingly, the artificial
earthquake had no effect on the accelerometer located at M2. 

Finally, the accelerometer was clumped to the telescope pier inside
the dome (location 5). As expected, the acceleration was small (0.2,
0.2, 0.5 mV rms in X, Y, Z). During this record, a minor natural
tremor occurred, causing some oscillation around 20\,Hz. Apart from
the tremor, only the 50 Hz line was detected.


Resonances at the  M2 unit were probed on November 12, 2015. The acceleration
was recorded  during 464\,s  while I tapped  at various  elements around
the secondary  mirror M2. The laser  launch telescope (LLT)  is attached
behind the  M2 unit, which is  connected to the telescope  top ring by
four spiders. The M2 is surrounded by a conic baffle. The telescope
was in horizontal position, the X axis of the accelerometer was
oriented horizontally. 

Figure~\ref{fig:top} shows the full-length  record in the signal in X
with indication of time and place of the artificial excitation. In all
cases, tapping  excited a resonance at  14.4\,Hz in the X  and Y axes,
and   this   quasi-sinusoidal    oscillation   decayed   approximately
exponentially ($1/e$ decrement of  6--10\,s). The accelerations in
X and  Y were  perfectly correlated  and of  similar amplitude,  so the
vibration was directed  at 45$^\circ$ relative to  the spiders.  There
was no  phase shift  (the XY  trajectory was  linear), and  the 14.4\,Hz
resonance was not detected in the Z axis.  A weaker resonance at 51.5\,Hz
was seen when the LLT or vertical spider were tapped.

In  summary, the  accelerometer tests  did not  reveal any  structural
resonances  that  could  explain  the   amplification  of  the  50  Hz
vibration. However, the resonances of  the M1 support structure around
66\,Hz,   likely  associated   with   the   wavefront  oscillation   in
astigmatism, were documented.


\section{Study of the Tip-Tilt Mirror M3}
\label{sec:M3}  


\begin{figure}
\begin{center}
\begin{tabular}{c}
\includegraphics[width=10cm]{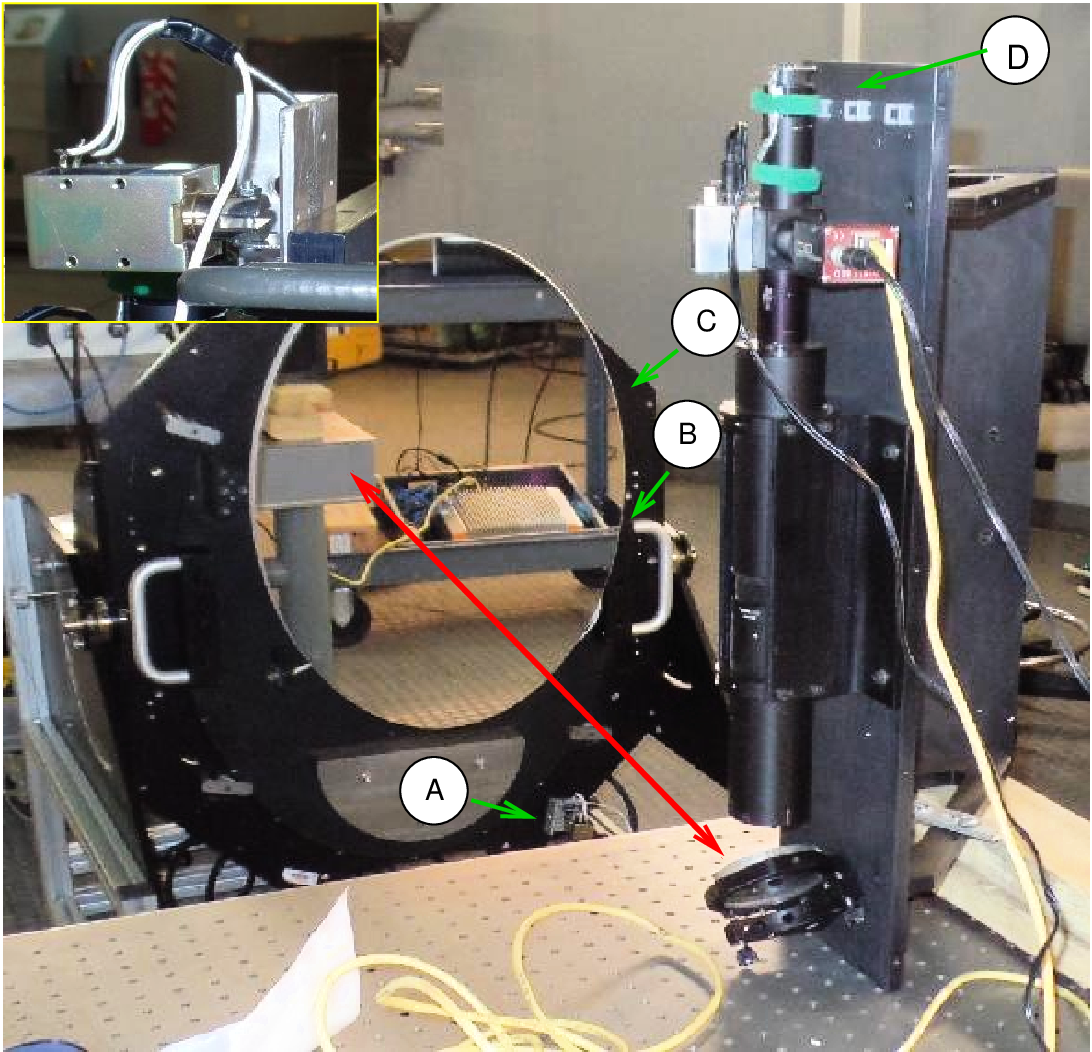}
\end{tabular}
\end{center}
\caption{Configuration  of  the M3  optical  test  on 2018-11-23.  The
  optical path is shown in red.   The insert on the top-left shows the
  shaker. Letters A--D indicate locations where the shaker was clumped.
\label{fig:M3test} }
\end{figure} 

The SOAR fast tip-tilt mirror, M3,  is suspended on a 2-axis pivot and
controlled  by the  digital servo  system that  replaced the  original
analog circuitry.  \cite{Warner2010} The  mirror tilts relative to its
base  are measured  by four  inductive sensors.   The servo  system is
designed to  damp the natural  resonance of  the M3 suspension  and to
provide a  bandwidth up  to 50\,Hz.  Potentially,  the M3  servo could
amplify the 50 Hz frequency because it is near the limit of its working band.

During  the  SOAR coating  shutdown  in  2018  November, the  M3  
response to mechanical  excitation at 50\,Hz was tested.   The M3 unit
was mounted in its test rig located  on the dome floor, with its servo
system active as  in normal operation.  The mirror angles were  measured by a
collimated laser beam emitted by the dome-probe device used previously
for  monitoring seeing  inside telescope  enclosure \cite{Bustos2018}.
It  produces a  4-cm  diameter  beam.  The  beam  is  directed to  M3,
reflected back,  and detected  by the  CCD camera  in the  same device
(Fig.~\ref{fig:M3test}).   The  dome probe  is  fixed  to the  optical
table,  and a  45$^\circ$ flat  folding  mirror in  a kinematic  mount
allows centering  the image  of the  reflected beam  on the  CCD.  The
software written  in python  enables fast acquisition  of   the spot
images, measurement  of  centroids, and  their statistical analysis
(rms,  power  spectrum).  Sequences  of  centroids  can be  saved  for
further study.  In these tests,  we used  the exposure time  of 5\,ms
(frame rate 200 Hz), sampling the 50-Hz signal adequately.

Mechanical vibrations with  a frequency of 50 Hz were  generated by an
improvised shaker clumped to the  M3 unit at different locations. The
shaker  is based  on  the Ledex  latching  solenoid (model  TDS-K12SB,
DC-12V).  The core of the solenoid is retained by the permanent magnet
and is released  when the voltage of $\sim$10\,V  of adequate polarity
is applied to  cancel the magnetic field.  The  solenoid was connected
to  the 50-Hz  AC voltage  with  adjustable amplitude  derived from  a
reducing transformer. The core was fixed to an aluminum bracket which,
in turn,  was clamped to the  structure.  The  coil  was connected to
the bracket flexibly,  allowing its motion.  The  core position inside
the  coil was  chosen to  obtain maximum  vibration.  Motion  of the
solenoid at 50 Hz creates a reaction  force applied to the base on the
order  of 1  newton for  a driving  voltage of  7--9\,V used  in these
tests.

\begin{figure}
\begin{center}
\begin{tabular}{c}
\includegraphics[width=16cm]{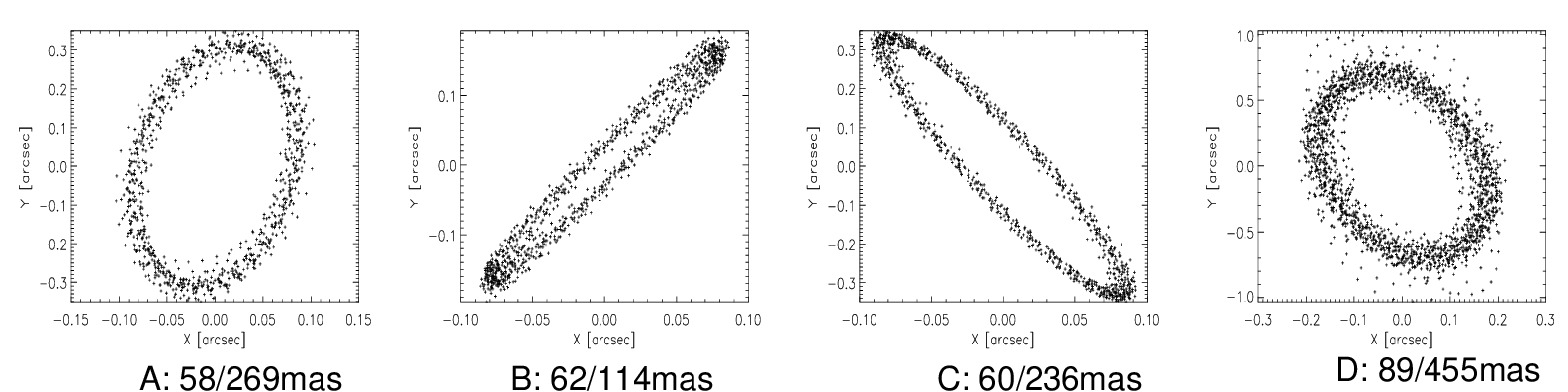}
\end{tabular}
\end{center}
\caption{Trajectory  of  the  M3   optical  axis  motion  with  50\,Hz
  frequency  excited by the  shaker  at locations  A-D.   The  numbers  indicate  rms
  amplitude of the cenrtroid motion in X and Y (in mas).
\label{fig:ellipses} }
\end{figure}

As both  the M3  cart and  the optical  table were  placed on  the dome
floor, any motion of persons inside  the dome affected the optics.  The
tests were  conducted on 2018-11-21  and 2018-11-23 without  people at
the  dome floor,  and  an rms  beam  motion as  small  as $0.1''$  was
achieved.  The  residual slow  mechanical perturbations caused  by the
floor instability did not affect our  tests.  Several records of the M3
internal  position  sensors  with   20\,kHz  sampling  were  taken  as
well. During  these tests, SOAR  was powered  by a generator,  and the
line frequency was 50.4\,Hz.

The 50-Hz  vibration excited by the  shaker is manifested by  a strong
peak in the power spectra of the dome probe signal; when the shaker is
switched  off, the  peak  disappears. Figure~\ref{fig:ellipses}  shows
representative trajectories of the reflected  beam in 10-s segments of
the   dome  probe   data  with   the  shaker   clumped  at   different
locations. The signal was filtered in the 47--53 Hz band.  The numbers
indicate the rms  centroid motion in X and Y  directions. These angles
are $\sim$15 times larger than the equivalent image motion at the SOAR
focus (M3 is  located at 4.2\,m before the focus,  while the effective
focal length of SOAR is 68\,m).   Obviously, the phase shift between X
and Y oscillations  and their amplitude depend on the  location of the
vibration   source.   One   might  question   these  results   because
narrow-band  filtering  of  noise  also  produces  a  quasi-sinusoidal
signal. However, when the shaker is turned off,  the XY
trajectory  resembles  a random  scatter with  an order  of magnitude
smaller amplitude  of 5--9\,mas.  When  the shaker is attached  to the
dome probe (location D), the 50-Hz vibration appears again.

When  the beam  trajectory  is computed  separately  for several  10-s
fragments  of  one dome  probe  signal  record, the  trajectories  and
amplitudes do not remain the  same, indicating that the M3 oscillation
is  excited  and damped  randomly.  However,  the orientation  of  the
ellipse, governed by the phase shift between X and Y, is approximately
preserved.

The 50-Hz oscillations excited by the  shaker are also apparent in the
signals of  the M3  position sensor recorded  during these  tests. The
trajectory  of the  optical  axis  deduced from  the  sensors is  also
elliptical.  An earlier record of the M3 position sensors was taken on
2014-07-08, when the  telescope was tracking.  The  M3 elevation angle
power spectrum had  a peak at 50\,Hz  (as well as smaller  peaks at 47
and 65\,Hz),  while the azimuth  angle had  a strong peak  at 100\,Hz.
The latter  could be of  electrical origin (feed-through of  the power
supply pulsation?).

\section{Mount Jitter}
\label{sec:jitter}  

For completeness, this section covers image oscillations produced by
tracking errors of the SOAR mount. Speckle interferometry
observations  collected a rich material for such study. The data cubes
are typically taken with  exposure time of 25\,ms  and
contain 400 or 600 individual images (10--15 s duration). Centroids of
the star in each frame are computed by the standard speckle pipeline
and saved. In most cases, the rms centroid motion in each speckle cube
is $\sim$0.3$''$. Its power spectrum is dominated by low frequencies,
as expected for image shifts produced by seeing and, occasionally,
wind shake. In some observing runs, however, quasi-periodic image
displacements caused by the mount tracking errors are present. 

A  large jitter  in  the  azimuth direction  has  been encountered  on
February  26-27,  2024.  It  appeared  and  disappeared  sporadically,
reaching a $2''$ amplitude in the worst cases. The jitter was stronger
at  low  telescope elevation.  The  reason  of  this jitter  has  been
identified: during the  daytime technical  work on the  mount, one  of the
tachometers of the azimuth motors  was connected wrongly. It should be
mentioned  that for  several years  SOAR operates  with three  azimuth
motors  instead  of  the  nominal four,  which  affects  the  backlash
compensation. Furthermore,  in 2026  another azimuth motor  failed and
problems  in the  motor controllers  were discovered.  These technical
faults directly affect the SOAR tracking quality.

\begin{figure}
\begin{center}
\begin{tabular}{c}
\includegraphics[width=12cm]{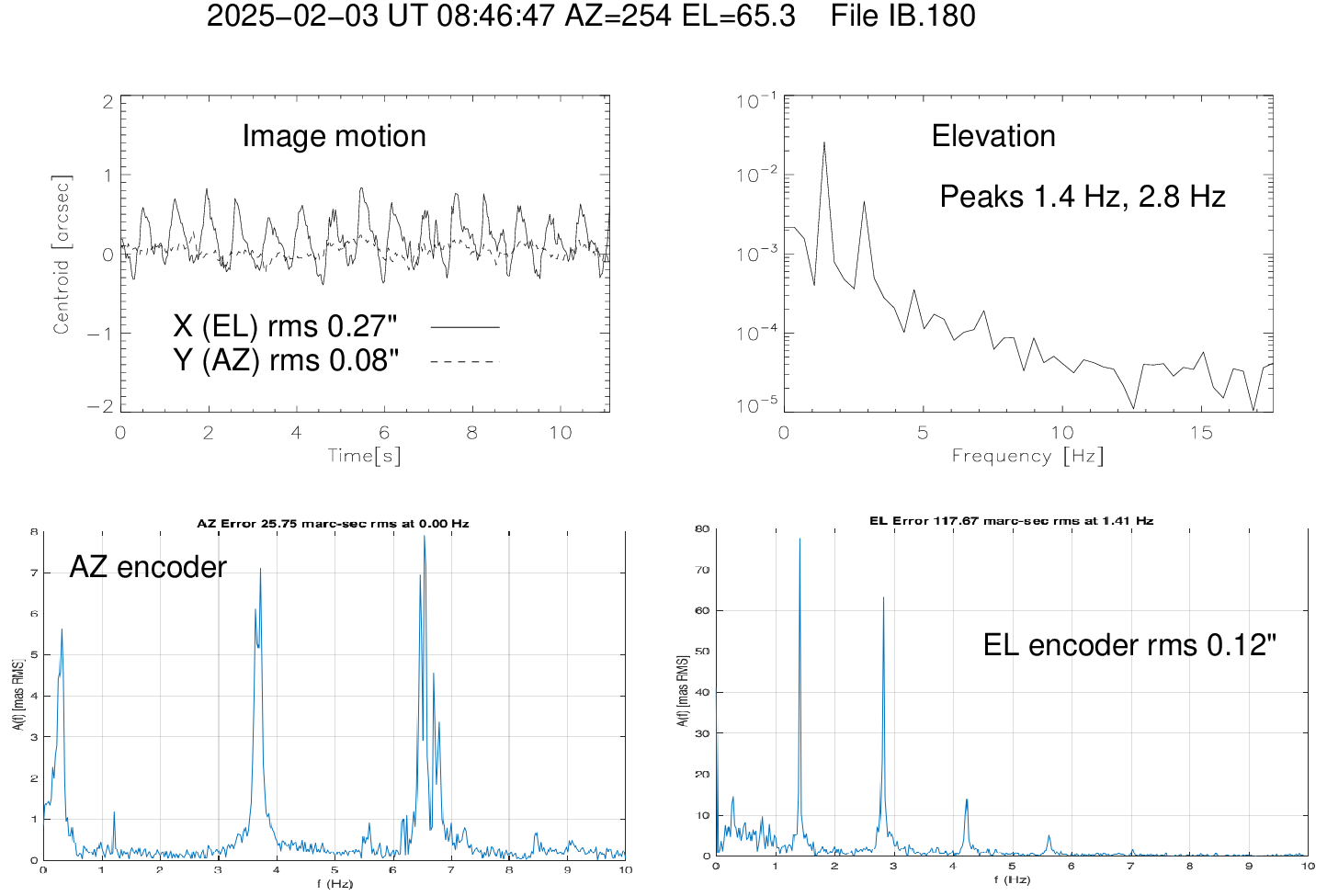}
\end{tabular}
\end{center}
\caption{Tracking errors of the SOAR mount deduced from the speckle data
  cubes. The upper panels show the trajectories of the image centroid
  in X and Y and the power spectrum of the X-motion (X is directed
  along  elevation). The lower panels show the power spectra
  of the mount jitter in azimuth and elevation derived from the
  encoder signals by B.~Cancino for the same time period.  
\label{fig:IB.180} }
\end{figure} 

\begin{figure}
\begin{center}
\begin{tabular}{c}
\includegraphics[width=12cm]{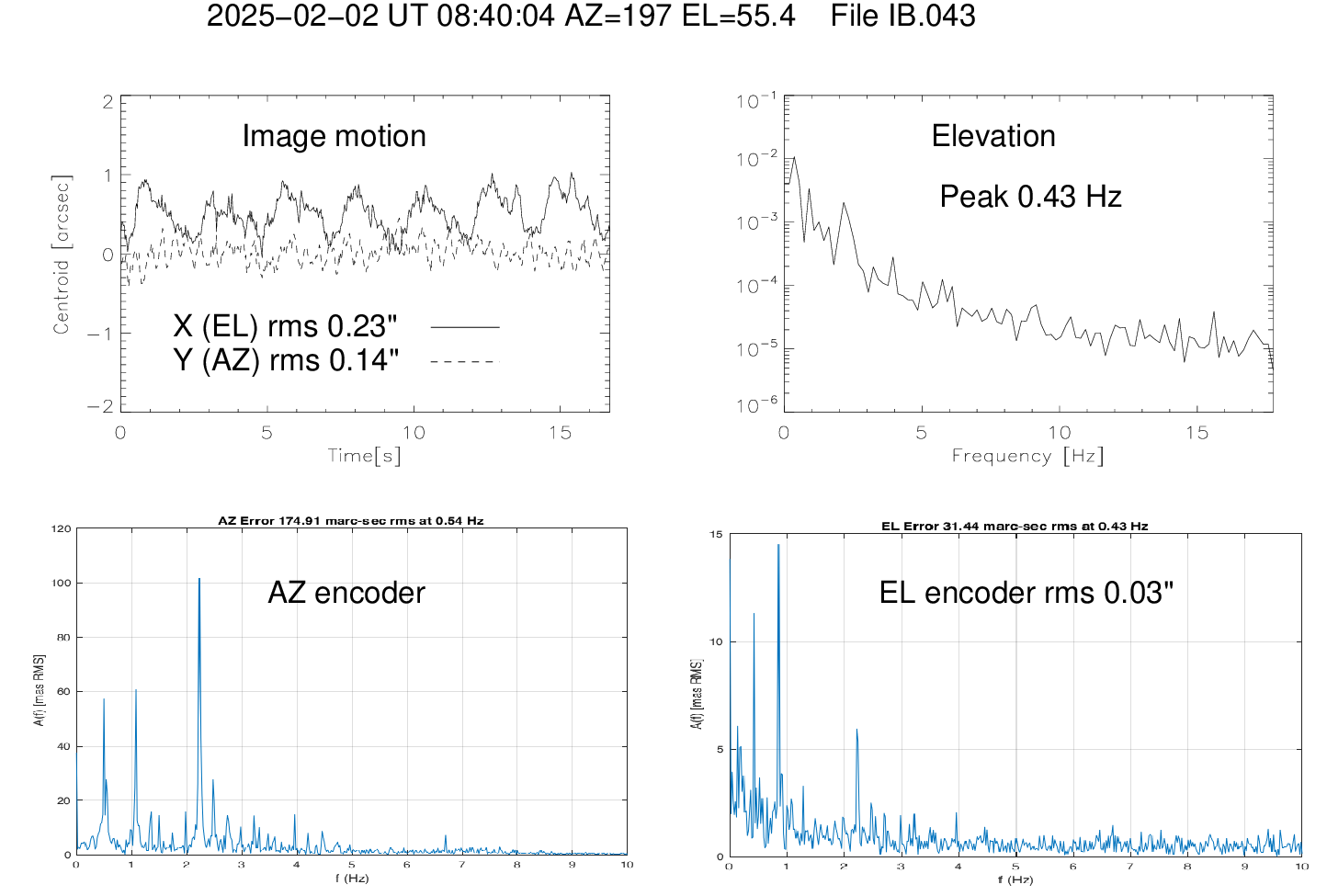}
\end{tabular}
\end{center}
\caption{Tracking errors of the SOAR mount deduced from the speckle data
  cubes (see Fig.~\ref{fig:IB.180}). 
\label{fig:IB.043} }
\end{figure} 

A more typical  situation was encountered on February  1-3, 2025, when
speckle  observations   were  conducted  in  the   morning  hours.   A
quasi-periodic oscillation  in the elevation direction  was noted. Its
amplitude  and  frequency  substantially  depended  on  the  telescope
pointing. Two  examples are presented in  Figures~\ref{fig:IB.180} and
\ref{fig:IB.043}.  In the  subsequent  analysis, shifts  of the  image
centroid were transformed (rotated) to elevation (X) and azimuth (Y)
directions.   Signals  of the  mount  encoders  recorded in  the  SOAR
telemetry system were examined by B.~Cancino as additional evidence.

When  the  telescope  points  east  or west,  the  tracking  speed  in
elevation is the largest. The frequency and amplitude of the elevation
oscillation  are also  larger. The  periodic jitter  is caused  by the
encoders.  The periods of encoder tapes are $8.904''$ and $20.24''$ in
elevation  and azimuth,  respectively.   The sine  and cosine  encoder
signals  are  interpolated  on  a  4096-point  grid,  and  any  signal
distortion caused by  an imperfect alignment of  the tape-reading head
translates into periodic position errors.   The bandwidth of the mount
servo  is  about  1\,Hz.   At  fast tracking,  the  first  and  second
harmonics  of the  periodic encoder  errors  are close to  or beyond  the
cutoff frequency of the mount servo, so they are amplified rather than
compensated.   In Fig.~\ref{fig:IB.180},  the rms  centroid motion  in
elevation is  $0.27''$, while the  rms tracking error measured  by the
elevation  encoders  is less,  $0.12''$.   In  contrast, the  centroid
motion in azimuth is only $0.08''$ rms, and the periodic errors in the
azimuth  encoders  are also  much  smaller.   Note that  during  these
observations the wind speed was low.

Figure~\ref{fig:IB.043} presents  similar data for the  case when SOAR
pointed in  the southern direction  and the tracking in  elevation was
slow.  The motion  in  elevation  is still  quasi-periodic  and has  a
comparable  amplitude  of  $0.23''$.  However, the  dominant  peak  at
0.43\,Hz is  now within  the mount servo  bandwidth, and  the tracking
servo faithfully  follows the  encoder signals. As  a result,  the rms
tracking error in elevation reported by the encoders is very small.

Although the  periodic tracking jitter  is slow compared to  the 25-ms
exposure  time used  in  speckle  interferometry,  it   causes  measurable
degradation  of  the  speckle  contrast.   A  harmonic  motion  as  in
Fig.~\ref{fig:IB.180} corresponds to the  maximum speed of $2.4''$ per
second  that blurs  the  speckles up to  60\,mas,  twice the  diffraction
limit.  The   actual  speckle   power  spectra  clearly show    this
attenuation in the direction parallel to the jitter.

Periodic tracking  errors are noted  sometimes by the  SOAR operators.
Typical observations  use fast guiders  to compensate both  the wind shake
and the  atmospheric  tilts.   The  guiders correct  the  relatively  slow
periodic tracking errors using the  actuated M3 mirror. Periodic tilts
in  M3 are  often perceived  by the  operators as a guider malfunction,
while in fact they demonstrate  the opposite: the guiding system works
well.   Without fast  guiding, periodic  tracking errors can  cause appreciable
image  elongation. Even  a moderate  jitter with  an rms  amplitude of
$\sigma = 0.3''$ causes peak-to-peak image displacement of $2 \sqrt{2}
\sigma = 0.85''$.  When the guiders  are not used or operate with long
exposures, stars in the SOAR  science images can be elongated by the mount
jitter.

On  September 10,  2025 the  elevation encoders  were tuned  and their
periodic errors  were reduced, diminishing also  the periodic tracking
oscillations.  This was  confirmed by the speckle data  of September 7
and 11,  2025, that show a  substantial reduction of the  jitter after
the encoder tuning.  On September 11,  the rms centroid motion in both
directions was about  $0.15''$.  However, tuning of  the encoders does
not  remain stable,  being affected  by seismic  tremors, so  it needs
regular attention by the telescope crew.

\section{Summary and Discussion}
\label{sec:sum}  


This paper relates the story of  the discovery and study of mysterious
50-Hz vibration of the SOAR optical axis. Detection of acceleration at
this frequency  even at the base  of the telescope pier  made it clear
that the driving source was external  to the telescope.  It could be a
transformer or some  synchronous motor.  One could  actually feel this
vibration  by touching  lightly  the  dome walls.   This  line in  the
acceleration power spectrum had a  width of 0.25\,Hz and its amplitude
was  not  stationary.   After  2020,   the  vibration  was  no  longer
perceptible   in   the   speckle-interferometric  data.    The   50-Hz
acceleration  itself was  too  small compared  to  the typical  wobble
amplitude  of the  optical axis  (20--30 mas).   This wobble  with its
characteristically  elliptical trajectory  was  produced  by the  fast
til-tilt  mirror  M3.   Its  servo  system  apparently  amplified  the
external perturbation,  causing phase-shifted motion  in X and  Y.  It
cannot be excluded that the  50-Hz excitation originated within the M3
unit itself (e.g.  in the motor driving its turret).   The true source
of the 50-Hz acceleration and the reason for its gradual disappearance
after 2020 still remain unidentified.

The Lowell Discovery Telescope (LDT)\footnote{
\url{https://lowell.edu/research/research-telescopes/lowell-discovery-telescope-ldt/}}
 is  a cousin of SOAR with similar
 thin primary  mirror, optical truss, and  mount.  However,
the LDT  has a larger secondary mirror  and   is not  equipped with
actuated tip-tilt mirror  (all instruments are located  on-axis at the
Cassegrain  focus).   The LDT  is  also  frequently used  for  speckle
interferometry,  but it  does not  suffer  from the  50-Hz wobble,  as
communicated by their speckle team.  This is yet another indirect evidence that
the 50-Hz wobble at SOAR was generated by its M3 unit.

Several experiments with accelerometers described here revealed
structural resonances of SOAR, e.g. the 16-Hz resonance of the M2 unit and the
65-Hz resonance of the M1 support. The latter is detected in
the astigmatism coefficient measured by the AO system in normal
operation. However, the amplitude of this vibration is too small to
affect the delivered resolution. The same is true for the tiny focus
oscillations at 47 Hz excited by vibration from fans.  The M2
structure had a mechanical resonace at 20\,Hz reported in
Ref.~\citenum{Warner2004} before addition of the LLT and its
electronics that shifted this resonance to 16\,Hz.

Periodic tracking  errors are common  to many telescopes, and  SOAR is
not an  exception.  A very strong  tracking jitter at SOAR  was caused
occasionally  by malfunction  of the  drive system  (motors and  their
controllers). Smaller  oscillations with  frequencies of 0.2-2.5  Hz are
encountered  routinely due to   periodic  errors of  the
encoders. The frequency of these oscillations is directly proportional
to the  tracking speed, while  their amplitude depends  sensitively on
the  alignment  of  the  encoder heads.  The  alignment  is  regularly
perturbed  by tremors,  requiring careful monitoring  and/or a
better mechanical design of the encoder support. Tracking oscillations
are typically corrected by the fast guiders; otherwise, they can cause
noticeable elongation of long-exposure images. These oscillations also
affect the quality of speckle data by smearing speckles in one
direction.

The  methods outlined  here  to characterize  SOAR vibrations,  namely
analysis of  AO real-time  data and accelerometry,  can be  applied to
other telescopes. Excitation of  structural resonances by ``artificial
earthquakes'' is  an efficient  way of their  identification. However,
the optical study  of the M3 unit relied on  a non-standard equipment.
To  our  knowledge, speckle-interferometric  data  were  used here  to
evaluate tracking for the first time.

The  diffraction-limited resolution  $\lambda/D$  delivered  by AO  or
speckle  interferometry   becomes  finer  with   increasing  telescope
diameter $D$,  placing stronger  requirements on  vibration mitigation
and  tip-tilt  errors. The  use  of  LGS  does  not help  in  tip-tilt
compensation which relies on natural  and necessarily faint
  guide stars.  AO  systems at  large  telescopes are  normally
designed for correction  of atmospheric tilts, and  fast vibrations of
optical  axis may  undermine  their performance.   In principle,  fast
motions  can   be  sensed  by  accelerometers   or  seismometers  (see
Ref.~\citenum{Tok2000}    and    references   therein),    alleviating
requirements on guide stars. This idea  has been advocated a long time
ago but, so far, has not  been realized.  The lesson learned from SOAR
is that telescopes do not vibrate as solid bodies, so inertial sensors
are of little help for tilt correction.



\subsection* {Acknowledgments}
These studies were conducted during  several years. They relied on essential
help from the SOAR daytime crew and on creative inputs of engineers at
CTIO.  I   am  particularly   grateful  to  Michael   Warner,  Rolando
Cantarutti,  and  Braulio Cancino.   Based  on  data obtained  at  the
Southern  Astrophysical Research  (SOAR) telescope,  which is  a joint
project of the Minist\'erio da Ci\^encia, Tecnologia e Inova\c{c}\~{o}es do Brasil
(MCTI/LNA),  the   US  National  Science  Foundation’s   NOIRLab,  the
University of North Carolina at  Chapel Hill (UNC), and Michigan State
University (MSU).


\newcommand{\aj}{Astronomical Journal}
\newcommand{\actaa}{Acta Astronomica}
\newcommand{\araa}{Annual Review of Astron and Astrophys}
\newcommand{\apj}{Astrophysical Journal}
\newcommand{\apjl}{Astrophysical Journal, Letters}
\newcommand{\apjs}{Astrophysical Journal, Supplement}
\newcommand{\ao}{Applied Optics}
\newcommand{\apss}{Astrophysics and Space Science}
\newcommand{\aap}{Astronomy and Astrophysics}
\newcommand{\aapr}{Astronomy and Astrophysics Reviews}
\newcommand{\aaps}{Astronomy and Astrophysics, Supplement}
\newcommand{\azh}{Astronomicheskii Zhurnal}
\newcommand{\baas}{Bulletin of the AAS}
\newcommand{\caa}{Chinese Astronomy and Astrophysics}
\newcommand{\cjaa}{Chinese Journal of Astronomy and Astrophysics}
\newcommand{\icarus}{Icarus}
\newcommand{\jcap}{Journal of Cosmology and Astroparticle Physics}
\newcommand{\jrasc}{Journal of the RAS of Canada}
\newcommand{\memras}{Memoirs of the RAS}
\newcommand{\mnras}{Monthly Notices of the RAS}
\newcommand{\na}{New Astronomy}
\newcommand{\nar}{New Astronomy Review}
\newcommand{\pra}{Physical Review A: General Physics}
\newcommand{\prb}{Physical Review B: Solid State}
\newcommand{\prc}{Physical Review C}
\newcommand{\prd}{Physical Review D}
\newcommand{\pre}{Physical Review E}
\newcommand{\prl}{Physical Review Letters}
\newcommand{\pasa}{Publications of the Astron. Soc. of Australia}
\newcommand{\pasp}{Publications of the ASP}
\newcommand{\pasj}{Publications of the ASJ}
\newcommand{\rmxaa}{Revista Mexicana de Astronomia y Astrofisica}
\newcommand{\qjras}{Quarterly Journal of the RAS}
\newcommand{\skytel}{Sky and Telescope}
\newcommand{\solphys}{Solar Physics}
\newcommand{\sovast}{Soviet Astronomy}
\newcommand{\ssr}{Space Science Reviews}
\newcommand{\zap}{Zeitschrift fuer Astrophysik}
\newcommand{\nat}{Nature}
\newcommand{\iaucirc}{IAU Cirulars}
\newcommand{\aplett}{Astrophysics Letters}
\newcommand{\apspr}{Astrophysics Space Physics Research}
\newcommand{\bain}{Bulletin Astronomical Institute of the Netherlands}
\newcommand{\fcp}{Fundamental Cosmic Physics}
\newcommand{\gca}{Geochimica Cosmochimica Acta}
\newcommand{\grl}{Geophysics Research Letters}
\newcommand{\jcp}{Journal of Chemical Physics}
\newcommand{\jgr}{Journal of Geophysics Research}
\newcommand{\jqsrt}{Journal of Quantitiative Spectroscopy and Radiative Transfer}
\newcommand{\memsai}{Mem. Societa Astronomica Italiana}
\newcommand{\nphysa}{Nuclear Physics A}
\newcommand{\physrep}{Physics Reports}
\newcommand{\physscr}{Physica Scripta}
\newcommand{\planss}{Planetary Space Science}
\newcommand{\procspie}{Proceedings of the SPIE}

\bibliography{vibr.bib}   
\bibliographystyle{spiejour}   



\end{document}